\documentstyle[prl,aps,floats,epsfig]{revtex}

\begin{document}
\twocolumn[\hsize\textwidth\columnwidth\hsize\csname @twocolumnfalse\endcsname

\title{Systematic Vertex Corrections through Iterative Solution of Hedin's
Equations beyond the $GW$ Approximation}
\author{Arno Schindlmayr$^{1,*}$ and R. W. Godby$^{2,\dagger}$}
\address{$^1$Cavendish Laboratory, University of Cambridge, Madingley Road,
Cambridge CB3 0HE, United Kingdom}
\address{$^2$Department of Physics, University of York, Heslington,
York YO1 5DD, United Kingdom}
\date{Received 15 August 1997}
\maketitle

\begin{abstract}
We present a general procedure for obtaining progressively more accurate
functional expressions for the electron self-energy by iterative solution
of Hedin's coupled equations. The iterative process starting from Hartree
theory, which gives rise to the $GW$ approximation, is continued further,
and an explicit formula for the vertex function from the second full cycle
is given. Calculated excitation energies for a Hubbard Hamiltonian
demonstrate the convergence of the iterative process and provide further
strong justification for the $GW$ approximation.
\end{abstract}

]
\narrowtext

Much of the modern theory of many-body effects in the electronic structure
of solids relies on a closed set of coupled integral equations known as
Hedin's equations \cite{Hed65}, which connect the Green's function $G$ of a
system of interacting electrons with the self-energy operator $\Sigma$, the
polarization propagator $P$, the dynamically screened Coulomb interaction
$W$, and a vertex function $\Gamma$. Simultaneously solving Hedin's
equations for a specified external potential in principle yields the exact
Green's function and quasiparticle excitation spectrum without the need of
actually calculating the many-electron wavefunction. Unfortunately,
however, the relation between the quantities is not just purely numerical
but involves nontrivial functional derivatives, so that an automated
numerical solution is not feasible and approximate functional expressions
with simpler dependences have to be considered instead. Most calculations
of quasiparticle excitations in real materials employ the $GW$
approximation \cite{Hed65}, which uses intermediate operators from the
first cycle of an iterative solution of Hedin's equations starting from
Hartree theory as a zeroth order approximation. The $GW$ approximation
neglects diagrammatic vertex corrections both in the polarization
propagator and the self-energy. Its theoretical foundation lies in the
assumption that sufficient convergence has been reached after the initial
cycle, but rigorous evidence has so far been prevented by the inherent
mathematical difficulties associated with a continuation of the iterative
process. Some corroboration stems from the surprisingly good agreement with
experimental spectra for a wide range of semiconductors \cite{Hyb85,God86}
and simple metals \cite{Nor87}, while more recent studies of transition
metals and their oxides highlighted deficiencies in the calculated spectra
that clearly indicate a lack of convergence for materials with strong
electronic correlation \cite{Ary92}. The $GW$ approximation has since often
been reinterpreted as the first order term in an expansion of the exact
self-energy in terms of the screened Coulomb interaction, and considerable
effort has been spent to include second order contributions. However, such
attempts have failed to produce a general improvement in numerical accuracy
due to far-reaching cancellation between the additional terms \cite{Bob94}.
In this Letter, we return to the original spirit of solving Hedin's
equations iteratively and present a scheme that generates progressively
accurate functional expressions at arbitrary levels of iteration. Starting
from Hartree theory, we obtain an explicit formula for the vertex function
from the second iterative cycle, which mixes certain diagrams of different
orders in the screened interaction. As exemplified for a Hubbard
Hamiltonian, this approach indeed yields convergent excitation energies
beyond the $GW$ approximation.

In our notation, the initial zeroth order self-energy and corresponding
Green's function are labelled $\Sigma^{(0)}$ and $G^{(0)}$, respectively.
The sequence in which the operators are obtained during the first iterative
cycle then starts with $\Gamma^{(1)}$, followed by $P^{(1)}$, $W^{(1)}$,
$\Sigma^{(1)}$, and finally $G^{(1)}$. For two reasons the principal
mathematical difficulty in continuing this process towards convergence lies
in the calculation of the vertex function. First, it is defined only
implicitly through the Bethe-Salpeter equation
\begin{eqnarray} \label{bethe-salpeter}
\Gamma^{(n+1)}(1,2;3)
&=& \delta(1,2) \delta(1,3) + \int\!\! \frac{\delta
\Sigma^{(n)}(1,2)}{\delta G^{(n)}(4,5)} G^{(n)}(4,6) \nonumber \\
&&\times G^{(n)}(7,5) \Gamma^{(n+1)}(6,7;3) \,d(4,5,6,7)
\end{eqnarray}
with the labels 1,2,\ldots\ each denoting a set of position, time, and spin
variables. While integral equations for other operators are readily solved
by matrix inversion in Fourier space, the convolutions in
(\ref{bethe-salpeter}) cannot easily be disentangled, so that the
computational expense is prohibitive. Second, it contains the functional
derivative $\delta \Sigma^{(n)} / \delta G^{(n)}$, which is nontrivial
because the Green's function $G^{(n)}$ is not explicitly contained in
$\Sigma^{(n)}$ but only calculated from it by means of Dyson's equation
\begin{eqnarray} \label{dyson}
G^{(n)}(8,9) &=& G^{(0)}(8,9) + \int\!\! G^{(0)}(8,1) \nonumber \\
&&\times \Delta\Sigma^{(n)}(1,2) G^{(n)}(2,9) \,d(1,2) ,
\end{eqnarray}
where $\Delta\Sigma^{(n)} = \Sigma^{(n)} - \Sigma^{(0)}$. A numerical
treatment of the vertex function becomes feasible only if the functional
derivative is evaluated analytically, and if (\ref{bethe-salpeter}) can be
solved explicitly for $\Gamma^{(n+1)}$. In the following we present a
transformation that satisfies both requirements.

\begin{figure}
\epsfxsize=\columnwidth \epsfbox{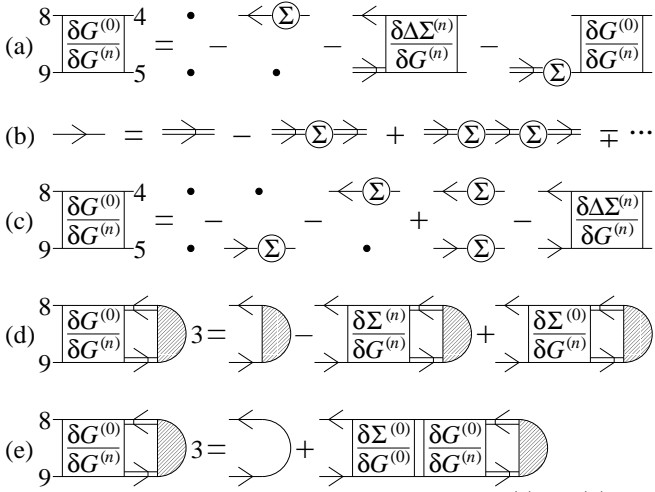}
\caption{The diagrammatic series (a) for $\delta G^{(0)} /
\delta G^{(n)}$ can be summed using an expansion of Dyson's equation in the
form (b) and has the explicit solution (c). By adding the first two terms
on the right-hand side of (d) and applying the chain rule to the third, we
obtain an integral equation (e) that can be solved by relation to the
Bethe-Salpeter equation for $n=0$. Single and double line arrows
indicate the Green's functions $G^{(0)}$ and $G^{(n)}$, respectively, the
encircled $\Sigma$ represents the self-energy correction
$\Delta\Sigma^{(n)} = \Sigma^{(n)} - \Sigma^{(0)}$ and the shaded
semicircle the vertex function $\Gamma^{(n+1)}$. Functional derivatives are
labelled explicitly.}
\label{fig:derivation}
\end{figure}

All operators relevant in this context contain as their basic building
block the zeroth order Green's function $G^{(0)}$. We therefore start by
applying the chain rule
\begin{equation} \label{sigman_over_gn}
\frac{\delta \Sigma^{(n)}(1,2)}{\delta G^{(n)}(4,5)}
= \int\!\! \frac{\delta \Sigma^{(n)}(1,2)}{\delta G^{(0)}(8,9)}
\frac{\delta G^{(0)}(8,9)}{\delta G^{(n)}(4,5)} \,d(8,9) .
\end{equation}
The first functional derivative can in principle be evaluated at any level
of iteration, so we focus on the second term. We perform the derivative of
(\ref{dyson}) with respect to $G^{(n)}$, which yields an integral equation
for the four-point operator $\delta G^{(0)} / \delta G^{(n)}$ as shown in
diagrammatic form in Fig.\ \ref{fig:derivation}(a). This series can be
summed using an expansion of Dyson's equation in terms of $G^{(n)}$ with
alternating signs [Fig.\ \ref{fig:derivation}(b)]. The explicit solution is
given in Fig.\ \ref{fig:derivation}(c). Employing Dyson's equation
once more, we can write this relation more concisely as
\begin{eqnarray}
\lefteqn{ \int\!\! \frac{\delta G^{(0)}(8,9)}{\delta G^{(n)}(4,5)}
G^{(n)}(4,6) G^{(n)}(7,5) \,d(4,5) } \nonumber \\
&=& G^{(0)}(8,6) G^{(0)}(7,9) - \int\!\! G^{(0)}(8,1) G^{(0)}(2,9)
\nonumber \\
&&\times \frac{\delta \Delta\Sigma^{(n)}(1,2)}{\delta G^{(n)}(4,5)}
G^{(n)}(4,6) G^{(n)}(7,5) \,d(1,2,4,5) ,
\end{eqnarray}
which still contains functional derivatives of the self-energies. Next we
resubstitute $\Delta\Sigma^{(n)} = \Sigma^{(n)} - \Sigma^{(0)}$ and perform
the integral with the vertex function $\Gamma^{(n+1)}$ as required by the
Bethe-Salpeter equation. These two steps yield the equation shown
diagrammatically in Fig.\ \ref{fig:derivation}(d). We simplify the sum of
the first two terms on the right-hand side using the relation
(\ref{bethe-salpeter}) and rewrite the functional derivative in the third
by applying the chain rule
\begin{equation}
\frac{\delta \Sigma^{(0)}(1,2)}{\delta G^{(n)}(4,5)}
= \int\!\! \frac{\delta \Sigma^{(0)}(1,2)}{\delta G^{(0)}(8,9)}
\frac{\delta G^{(0)}(8,9)}{\delta G^{(n)}(4,5)} \,d(8,9) .
\end{equation}
In this way we finally obtain the integral equation shown in Fig.\
\ref{fig:derivation}(e), which is closely related to the Bethe-Salpeter
equation for $n = 0$. By comparison, we find
\begin{eqnarray} \label{cancellation}
\lefteqn{ \int\!\! G^{(0)}(8,6) G^{(0)}(7,9) \Gamma^{(1)}(6,7;3) \,d(6,7) }
\nonumber \\
&=& \int\!\! \frac{\delta G^{(0)}(8,9)}{\delta G^{(n)}(4,5)} G^{(n)}(4,6)
G^{(n)}(7,5) \Gamma^{(n+1)}(6,7;3) \nonumber \\
&&\times \,d(4,5,6,7) .
\end{eqnarray}
We can now use this identity to sum the diagrammatic series in
(\ref{bethe-salpeter}) and rewrite it in the alternative form
\begin{eqnarray} \label{vertexfunction}
\Gamma^{(n+1)}(1,2;3)
&=& \delta(1,2) \delta(1,3) + \int\!\! \frac{\delta
\Sigma^{(n)}(1,2)}{\delta G^{(0)}(4,5)} G^{(0)}(4,6) \nonumber \\
&&\times G^{(0)}(7,5) \Gamma^{(1)}(6,7;3) \,d(4,5,6,7) .
\end{eqnarray}

This expression is remarkably similar to the original integral equation,
except that all operators but $\Sigma^{(n)}$ on the right-hand side are
replaced by their lowest order equivalents. While successive iterations
dress each propagator with new sets of diagrams, the identity
(\ref{cancellation}) indicates far-reaching cancellation between the
expansion terms from individual propagators, yielding a much simpler
expression for the vertex corrections than originally anticipated. The
emergence of propagators from the lowest iterative cycle reduces the
numerical expense substantially, as in practice mean-field theories are
used as a zeroth order approximation. In this case $G^{(0)}$ contains no
satellite spectrum but only a set of robust quasiparticle excitations. The
transformation also satisfies our requirements for a numerical treatment by
giving an explicit definition for the vertex function and expressing the
functional derivative in a way that can be evaluated at higher orders. In
the self-consistency limit $n \to \infty$, (\ref{vertexfunction}) implies a
relation between the exact self-energy and vertex function.

\begin{figure}
\epsfxsize=\columnwidth \epsfbox{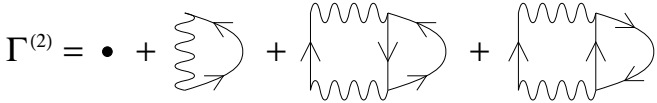}
\caption{Vertex correction $\Gamma^{(2)}$ to the $GW$ self-energy from the
second iterative cycle of Hedin's equations. Single line arrows and wiggly
lines represent the zeroth order Green's function $G^{(0)}$ and the
screened Coulomb interaction $W^{(1)}$ in the random phase approximation,
respectively.}
\label{fig:gamma2}
\end{figure}

In condensed matter physics, iteration conventionally starts from Hartree
theory as a zeroth order approximation, so that $\Sigma^{(0)} = 0$ and
$G^{(0)} = G^{\rm H}$. Due to the vanishing self-energy the functional
derivative in (\ref{bethe-salpeter}) is identically zero, yielding a
trivial vertex function. The subsequent iteration generates the $GW$
approximation
\begin{eqnarray}
P^{(1)}(1,2) &=& -i G^{(0)}(1,2) G^{(0)}(2,1) \label{p_rpa} \\
W^{(1)}(1,2) &=& v(1,2) + \int\!\! W^{(1)}(1,3) \nonumber \\
&&\times P^{(1)}(3,4) v(4,2) \,d(3,4) \\
\Sigma^{(1)}(1,2) &=& i G^{(0)}(1,2) W^{(1)}(1^+,2) , \label{sigma_gw}
\end{eqnarray}
where in the last equation 1$^+$ implies that a positive infinitesimal is
added to the time variable. $v$ is the bare Coulomb interaction. At
this level the self-energy is modelled as the product of the Green's
function $G^{(0)}$ and the screened interaction $W^{(1)}$ in the random
phase approximation (RPA). The functional form is reminiscent of the Fock
potential, but the electronic exchange includes dynamic screening and so
reaches beyond the limits of mean-field theories. Using this definition of
the self-energy, it is easy to verify that its derivative with respect to
the zeroth order Green's function is given by
\begin{eqnarray} \label{sigma1_over_g0}
\frac{\delta \Sigma^{(1)}(1,2)}{\delta G^{(0)}(4,5)}
&=& i \delta(1,4) \delta(2,5) W^{(1)}(1,2)  \nonumber \\
&&+ G^{(0)}(1,2) \left[ W^{(1)}(1,5) W^{(1)}(4,2) \right. \nonumber \\
&&+ \left. W^{(1)}(1,4) W^{(1)}(5,2) \right] G^{(0)}(5,4) .
\end{eqnarray}
The corresponding vertex function $\Gamma^{(2)}$ that underlies the second
iterative cycle is shown in diagrammatic form in Fig.\ \ref{fig:gamma2}.
This finite set of vertex corrections is distinct from that obtained
through expansion by orders of the screened interaction in that it
comprises selected terms which are of zeroth, first, and second order in
$W^{(1)}$.

Although the self-energy (\ref{sigma_gw}) was obtained by iteration
starting from Hartree theory, in practice it is more often evaluated using
a zeroth order Green's function $G^{(0)} = G^{\rm DFT}$ from a previous
density-functional calculation with  $\Sigma^{(0)} = V^{\rm XC}$ equal to
the exchange-correlation potential. Although this substitution violates the
original spirit of the iterative scheme, physical arguments suggest only a
small deviation to the propagators properly derived from density-functional
theory as a zeroth order approximation \cite{Hyb85}. Since the form of
$\Sigma^{(1)}$ remains identical to that considered before, its derivative
is still given by (\ref{sigma1_over_g0}). The nontrivial first order vertex
$\Gamma^{(1)}$ required to evaluate (\ref{vertexfunction}) was derived in
Ref.\ \cite{Sol94} and indeed found to have insignificant numerical impact
on the band gap of silicon. It is probably also negligible in the second
iteration of such a calculation, so that the expression shown in Fig.\
\ref{fig:gamma2} may still be used for $\Gamma^{(2)}$.

In the following we will supplement our discussion with the numerical
investigation of a system of strongly correlated electrons that explores
the effects of the vertex correction $\Gamma^{(2)}$. As the evaluation of
higher order diagrams for real materials would demand computational
resources beyond the scope of the present study, we consider a
two-dimensional square array of 3$\times$3 atoms described by the Hubbard
Hamiltonian
\begin{equation}
{\cal H} = \sum_{{\bf R},\sigma} \epsilon_{\bf R} \hat{n}_{{\bf R}\sigma}
-t \sum_{\langle{\bf R},{\bf R}'\rangle,\sigma} c^\dag_{{\bf R}\sigma}
c_{{\bf R}'\sigma} + U \sum_{\bf R} \hat{n}_{{\bf R}\uparrow}
\hat{n}_{{\bf R}\downarrow} .
\end{equation}
The model captures the essential physical features of materials with strong
electronic correlation such as the transition metals, which are known to
be inadequately described in the $GW$ approximation \cite{Ary92}, but is
simple enough to allow the calculation of full optical spectra rather than
just individual matrix elements as in Refs.\ \cite{Bob94,Sol94}.
Here, $c^\dag_{{\bf R}\sigma}$ and $c_{{\bf R}\sigma}$ are the creation
and annihilation operators for an electron at site ${\bf R}$ with spin
$\sigma$, and we define $\hat{n}_{{\bf R}\sigma} = c^\dag_{{\bf R}\sigma}
c_{{\bf R}\sigma}$. The notation $\langle {\bf R},{\bf R}' \rangle$ implies
a sum over nearest neighbors only. This model was introduced in Ref.\
\cite{Ver95} to compare variations of the $GW$ approximation, and its
performance within the first iterative cycle of Hedin's equations is thus
well understood. The single band of the cluster can accomodate up to 18
electrons; we consider a system of 16 electrons. The high fractional
band filling resembles that of the $d$ orbitals in the late transition
metals. Although we use open boundary conditions, the on-site energies
$\epsilon_{\bf R}$ are chosen in such a way as to yield uniform occupation
numbers in the Hartree approximation, as expected in infinite systems. For
reference, the on-site energy is $2t$ for corner sites, $t$ on edge sites,
and $0$ in the center of the cluster. We use medium interaction $U = 4t$
and set $t = 1$.

\begin{figure}
\epsfxsize=\columnwidth \epsfbox{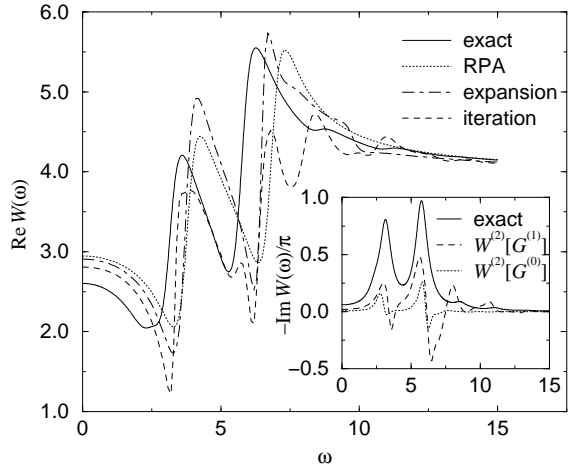}
\caption{The exact screened interaction compared to the RPA, an expansion
of the polarization to first order in $W$, and the second iteration of
Hedin's equations. The latter yields more accurate plasmon energies and a
qualitatively correct satellite spectrum. Inset: The shift in plasmon
energies is due to the explicit vertex function and remains when the
diagrams are evaluated with $G^{(0)}$ rather than $G^{(1)}$, while the
satellite peaks result from internal dressed propagators.}
\label{fig:plasmons}
\end{figure}

In Fig.\ \ref{fig:plasmons} we compare the exact screened interaction $W$
with its RPA counterpart from the first iteration of Hedin's equations, in
which the polarization propagator takes the form (\ref{p_rpa}), and with
the second iteration. Here
\begin{equation} \label{p_2nd}
P^{(2)}(1,2) = -i \int\!\! G^{(1)}(2,3) G^{(1)}(4,2) \Gamma^{(2)}(3,4;1)
\,d(3,4) .
\end{equation}
As in Ref.\ \cite{Ver95}, $G^{(1)}$ was obtained from a shifted zeroth
order Green's function in order to avoid problems with differing chemical
potentials in the perturbation series. The figure shows the diagonal matrix
element for the central site of the cluster, which because of the chosen
geometry corresponds most closely to the screening in extended systems.
Other matrix elements exhibit similar behavior, and the displayed curves
are thus representative. For comparison, we also show results from an
expansion of the polarization propagator beyond the RPA to first order in
$W^{(1)}$. This approach is qualitatively distinct in that only the
vertex function becomes dressed, while in the iterative scheme both
$\Gamma$ and the internal propagators $G$ in (\ref{p_2nd}) are
simultaneously updated.

The exact screening is dominated by a pair of strong plasmon peaks at 3 and
just under 6 eV, but three satellites at higher energy can be identified.
While qualitatively acceptable, the RPA as a first approximation ignores
the satellite spectrum and places the plasmons too high by about 1 eV. The
latter deficiency is somewhat improved by the further expansion of the
polarization propagator, but the description of the satellites remains poor
and is not even qualitatively correct. This is in line with previous
observations of far-reaching cancellation between the additional terms
\cite{Bob94}.

In comparison, the second iteration is more effective in shifting the
plasmon energies, particularly for the lower peak, and also yields a better
low-frequency limit for the static dielectric function. Furthermore, we
observe the emergence of a satellite spectrum that is in good agreement
with the exact curve concerning the number and position of features. Their
exaggerated spectral weight can be traced to the satellites in the $GW$
Green's function, which for this model are overestimated by a similar
factor \cite{Ver95}. To isolate the effect of the vertex function, we have
also evaluated (\ref{p_2nd}) using $G^{(0)}$ rather than $G^{(1)}$. The
resultant dielectric function, displayed in the inset in Fig.\
\ref{fig:plasmons}, retains the full shift in plasmon energies but no
satellites, which are due to the internal dressed propagators.
Incidentally, the graph also shows part of the apparent plasmon spectral
weight to have moved across the real axis. This incorrect analytic
structure is an unfortunate but common consequence of nontrivial vertex
corrections and due to the occurrence of higher order poles in $P^{(2)}$.
The same effect is well documented for the expansion in terms of the
screened Coulomb interaction \cite{Min74}, where it is also observed in the
present calculation. However, we have confirmed that the integral of the
spectral function $-{\rm Im} W^{(2)}(\omega) / \pi$ over the positive
half-axis is correctly greater than zero for all diagonal matrix elements,
and as intervals of negative spectral weight are few and embedded between
features with correct analytic behavior, they do not dominate convolutions
in the later course of the iteration.

In summary, we have presented a scheme for the systematic construction of
vertex corrections by iterative solution of Hedin's coupled equations, and
we have given explicit formulae for the propagators beyond the $GW$
approximation. Numerical results for a model of strongly correlated
electrons indicate that this method not only yields improved excitation
energies but is also more powerful than a comparable expansion by orders of
the screened Coulomb interaction, in particular by generating a superior
satellite spectrum. On a fundamental level, these findings provide the
first direct evidence of convergence of the iterative approach and thus
give further theoretical justification for the $GW$ approximation.

This work was supported by the Engineering and Physical Sciences Research
Council. One of the authors (A.S.) acknowledges additional funding from the
Deut\-scher Aka\-de\-mi\-scher Aus\-tausch\-dienst under its HSP III
scheme, the Stu\-di\-en\-stif\-tung des deut\-schen Vol\-kes, the
Gott\-lieb Daimler- und Karl Benz-Stiftung, and Pembroke College Cambridge.

\end{document}